\documentclass[twocolumn,accepted=2019-08-24]{quantumarticle}

\pdfoutput=1

\usepackage[numbers,sort&compress]{natbib}
\usepackage{epsfig}
\usepackage{amsmath,amssymb,amsfonts}
\usepackage{mathrsfs}
\usepackage{bbm}
\usepackage{slashed}
\usepackage{graphicx}
\usepackage{verbatim}
\usepackage[usenames]{color}
\usepackage{mathtools}
\usepackage{algorithm}
\usepackage{algorithmicx}
\usepackage{algpseudocode}

\date{}
\begin{document}
	\title{Quantum error correction for the toric code using deep reinforcement learning }
	\author{Philip Andreasson}
	\author{Joel Johansson}
	\author{Simon Liljestrand}
	\author{Mats Granath}
	\email[]{mats.granath@physics.gu.se}
	\affiliation{Department of Physics, University of Gothenburg,
		SE-41296 Gothenburg, Sweden}
	
\begin{abstract}
We implement a quantum error correction algorithm for bit-flip errors on the topological toric code using deep reinforcement learning. An action-value Q-function encodes the discounted value of moving a defect to a neighboring site on the square grid (the action) depending on the full set of defects on the torus (the syndrome or state). The Q-function is represented by a deep convolutional neural network. Using the translational invariance on the torus allows for viewing each defect from a central perspective which significantly simplifies the state space representation independently of the number of defect pairs. The training is done using experience replay, where data from the algorithm being played out is stored and used for mini-batch upgrade of the Q-network. We find performance which is close to, and for small error rates asymptotically equivalent to, that achieved by the Minimum Weight Perfect Matching algorithm for code distances up to $d=7$. Our results show that it is possible for a self-trained agent without supervision or support algorithms to find a decoding scheme that performs on par with hand-made algorithms, opening up for future machine engineered decoders for more general error models and error correcting codes.

\end{abstract}
%
%
\maketitle
\section{Introduction}
Much of the spectacular advances in machine learning using artificial neural networks has been in the domain of supervised learning were deep convolutional networks excel at categorizing objects when trained with big annotated data sets\cite{krizhevsky2012imagenet,lecun2015deep,goodfellow2016deep}. A different but also more challenging type of problem is when there is no a priori solution key, but rather a dynamic environment through which we want to learn to navigate for an optimal outcome. For these types of problems reinforcement learning (RL) \cite{sutton2018reinforcement} combined with deep learning has had great success recently when applied to problems such as computer and board games\cite{tesauro1995temporal,mnih2013playing,mnih2015human,silver2017mastering}. The super-human performance achieved by deep reinforcement learning has revolutionized the field of artificial intelligence and opens up for applications in many areas of science and technology.    

In physics the use of machine learning  has seen a great deal of interest lately\cite{arsenault2014machine,van2017learning,carrasquilla2017machine,carleo2017solving,gao2017efficient}. The most natural type of application of neural networks is in the form of supervised learning where the deep network can capture correlations or subtle information in real or artificial data. The use of deep reinforcement learning may be less obvious in general as the type of topics addressed by RL typically involve some sort of "intelligent" best strategy search, contrary to the deterministic or statistical models used in physics. 

In this paper we study a type of problem where artificial intelligence is applicable, namely the task of finding a best strategy for error correction of a topological quantum code; the potential basic building blocks of a quantum computer. In the field of quantum computing, smart algorithms are needed for error correction of fragile quantum bits\cite{PhysRevA.52.R2493,PhysRevLett.77.793,nielsen2002quantum,terhal2015quantum}. Reinforcement learning has been suggested recently as a tool for quantum error correction and quantum control\cite{melnikov2018active,PhysRevX.8.031084,PhysRevX.8.031086}, where an agent  learns to manipulate a quantum device that functions in an imperfect environment and with incomplete information. Under the umbrella term "Quantum Machine Learning" there are also interesting prospects of utilizing the natural parallelization of quantum computers for machine learning itself\cite{biamonte2017quantum}, but we will be dealing here with the task of putting (classical) deep learning and AI at the service of quantum computing. 

Due to the inherently fragile nature of quantum information a future universal quantum computer will require quantum error correction.\cite{PhysRevA.52.R2493,PhysRevLett.77.793,nielsen2002quantum,terhal2015quantum}
 Perhaps the most promising framework is to use topological error correcting codes.\cite{kitaev2003fault,dennis2002topological,raussendorf2007topological,fowler2012surface,kelly2015state}
Here, logical qubits consisting of a large set of entangled physical qubits are protected against local disturbances from phase or bit flip errors as logical operations require global changes.  
Local stabilizer operators, in the form of parity checks on a group of physical qubits, provide a quantum non-demolition diagnosis of the logical qubit in terms of violated stabilizers; the so-called syndrome. In order for errors not to proliferate and cause logical failure, a decoder, that provides a set of recovery operations for correction of errors given a particular syndrome, is required. As the syndrome does not uniquely determine the physical errors, the decoder has to incorporate the statistics of errors corresponding to any given syndrome. In addition the syndrome itself may be imperfect, due to stabilizer measurement errors, in which case the decoder must also take that into account.   

In the present work we consider Kitaev's toric code\cite{kitaev2003fault,dennis2002topological,fowler2012surface} which is a stabilizer code formulated on a square lattice with periodic boundary conditions (see Figure \ref{fig:model} and Section \ref{sec:toric}). 
We will only consider bit-flip errors which correspond to syndromes with one type of violated stabilizer that can be represented as plaquette defects on the lattice (see Figure \ref{fig:errorchain}). The standard decoder for the toric code is the Minimum Weight Perfect Matching (MWPM) or Blossom algorithm\cite{edmonds1965paths,fowler2015minimum,PhysRevA.90.032326} that works by finding the pairwise matching of syndrome defects with shortest total distance, corresponding to the minimal number of errors consistent with the syndrome. 
The decoder problem is also conceptually well suited for reinforcement learning, similar in spirit to a board game; the state of the system is given by the syndrome, actions correspond to moving defects of the syndrome, and with reward given depending on the outcome of the game. By playing the game, the agent improves its error correcting strategies and the decoder is the trained agent that provides step by step error correction. As in any RL problem the reward scheme is crucial for good performance. The size of the state-action space is also a challenge, to provide the best action for each of a myriad syndromes, but this is exactly the problem addressed by recent deep learning approaches to RL.\cite{tesauro1995temporal,mnih2013playing,mnih2015human,silver2017mastering} 

We find that by setting up a reward scheme that encourages the elimination of the syndrome in as few operations as possible within the deep Q-learning (or deep Q-network, DQN)\cite{mnih2013playing,mnih2015human} formalism we are able to train a decoder that is comparable in performance to MWPM. Although the present algorithm does not outperform the latter we expect that it has the potential to be more versatile when addressing depolarizing noise (with correlated bit and phase flip errors), measurement noise giving imperfect syndromes, or varying code geometries.  Compared to the MWPM algorithm the RL algorithm also has the possible advantage that it provides step by step correction whereas the MWPM algorithm only provides information on which defects should be paired, making the former more adaptable to the introduction of additional errors.


In concurrent work by \citet{sweke2018reinforcement} an application of reinforcement learning to error correction of the toric code was implemented. That work focuses on the important issue of imperfect syndromes as well as depolarizing noise and used an auxiliary "referee decoder" to assist the RL decoder. In the present work we consider the simpler but conceptually more direct problem of error correction on a perfect syndrome and with only bit flip error. Also in contrast to \cite{sweke2018reinforcement} we study the actual "toric" code, rather than the code with boundaries. Clearly the toric code will be harder to implement experimentally but nevertheless provides a well understood standard model. It also provides a simplification from the fact that on a torus only the relative positions of syndrome defects are relevant which reduces the state space complexity that decoder agent has to master. By focusing on this minimal problem we find that we can make a rigorous benchmark on the RL decoder showing near optimal performance.

Finding better performing decoders has been the topic of many studies, using methods such as renormalization group\cite{duclos2010fast,Duclos-Cianci:2014:FRG:2638670.2638671}, cellular automata\cite{herold2015cellular,kubica2018cellular}, and a number of neural network based decoders\cite{PhysRevLett.119.030501,PhysRevX.8.031084,krastanov2017deep,varsamopoulos2017decoding,baireuther2018machine,breuckmann2018scalable,chamberland2018deep,maskara2018advantages,ni2018neural,PhysRevLett.122.200501}. The decoder presented in this paper does not outperform state of the art decoders, it's value lies in showing that it is possible to use reinforcement learning to achieve excellent performance on a minimal model. Given that deep reinforcement learning is arguably the most promising AI framework it holds prospect for future versatile self-trained decoders that can adapt to different error scenarios and code architectures.

The outline of the paper is the following. In the {\em Background} section we give a brief but self-contained summary of the main features of the toric code including the basic structure of the error correction and a similar summary of one-step Q-learning and deep Q-learning. The following section, {\em RL Algorithm}, describes the formulation and training of the error correcting agent. In the {\em Results} section we shows that we have trained the RL agent up to code distance $d=7$ with performance which is very close to the MWPM algorithm. We finally conclude and append details of the asymptotic fail rate for small error rates as well as the neural network architecture and the RL and network hyperparameters.

\section{Background}
\subsection{Toric code\label{sec:toric}}

\begin{figure}
    \includegraphics[scale=0.6]{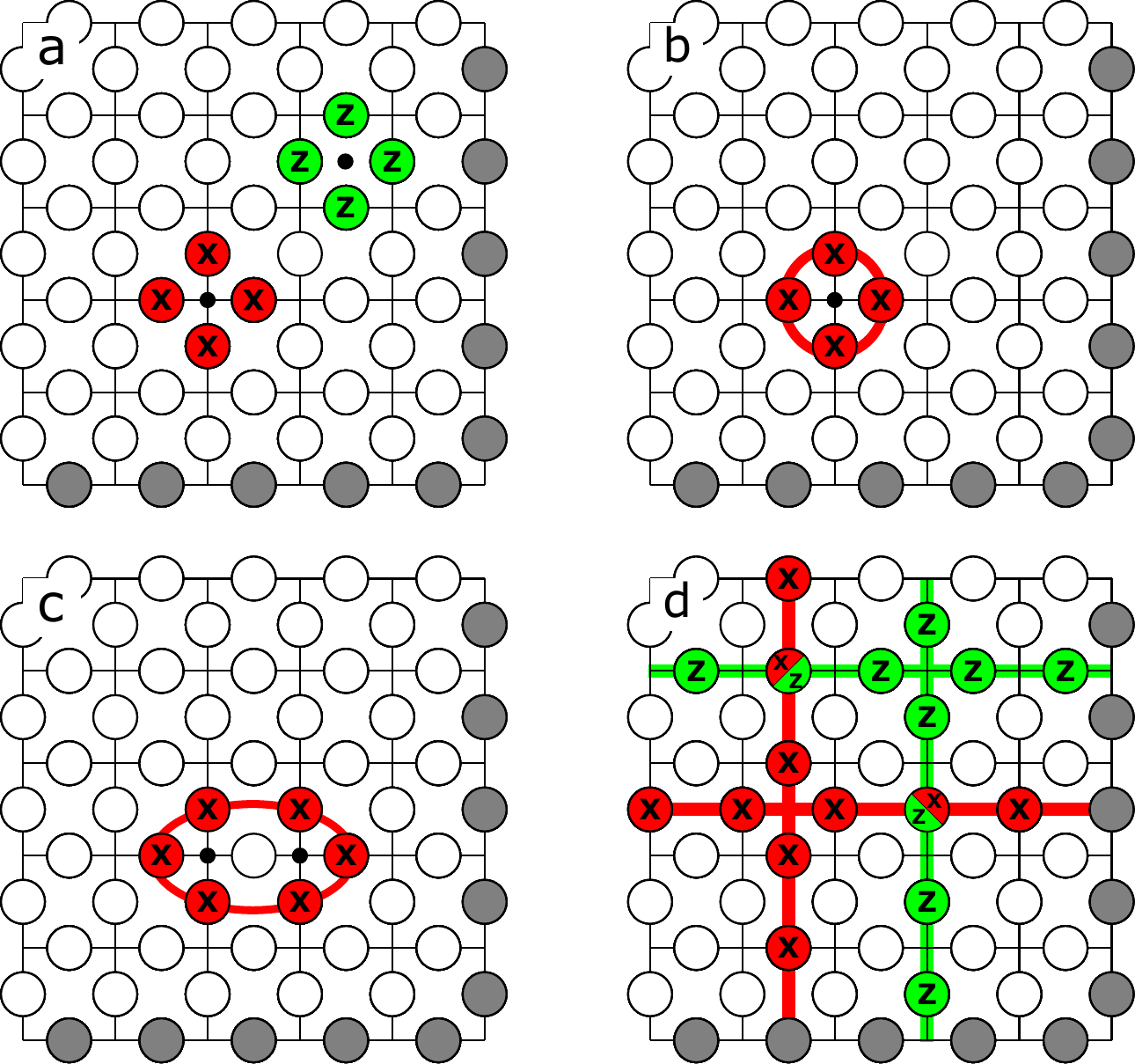} 
    \caption{\label{fig:model} A $d=5$ toric code with rings indicating the physical qubits and grey showing the periodic boundary conditions. a) Plaquette (green) and vertex (red) stabilizer operators, as products of $\sigma^z$ and $\sigma^x$ Pauli matrices. b) A single vertex operator can be represented as a loop flipping the qubits that the it crosses. c) Two neighboring vertex operators make up a larger loop. d) The logical operators $\bar{X}_{1/2}$ (red) and $\bar{Z}_{1/2}$ (green) consist of loops winding the torus and are not representable in terms of products of vertex or plaquette operators.}
\end{figure}

The basic construction of the toric code is a square lattice with a spin-$\frac{1}{2}$ degree of freedom on every bond, the physical qubits, and with periodic boundary conditions, as seen in Figure \ref{fig:model}.\cite{kitaev2003fault,dennis2002topological}\footnote{Figures in this section were inspired by lecture notes \cite{Browne.lecture}.} (An alternative rotated lattice representation with the qubits on sites is also common in the literature.)  The model is given in terms of a Hamiltonian 
\begin{equation}
    H=-\sum_{\alpha}\hat{P}_\alpha-\sum_{\nu}\hat{V}_{\nu}\,,
\end{equation}
where $\alpha$ runs over all plaquettes and $\nu$ over all vertices (sites). The stabilizers are the plaquette operators $\hat{P}_\alpha=\prod_{i\in\alpha}\sigma^z_i$ and the vertex operators $\hat{V}_\nu=\prod_{i\in\nu}\sigma^x_i$, where $\sigma^z$ and $\sigma^x$ are the Pauli matrices. (Where, in the $\sigma^z$ basis, $\sigma^z|\uparrow/\downarrow\rangle=\pm 1 |\uparrow/\downarrow\rangle$ and $\sigma^x|\uparrow/\downarrow\rangle=|\downarrow/\uparrow\rangle$.) The stabilizers  commute with each other and the Hamiltonian thus block diagonalizing the latter. On a $d\times d$ lattice of plaquettes $d^2-1$ plaquette operators are linearly independent (e.g. it is not possible to have a single $-1$ eigenvalue with all other $+1$) and correspondingly for the vertex operators. With $2d^2$ physical qubits and $2d^2-2$ stabilizers the size of each block is $2^{2d^2}/2^{2d^2-2}=4$, corresponding in particular to a ground state which is 4-fold degenerate. These are the states that will serve as the logical qubits. (More precisely, given the 4-fold degeneracy it is a qudit or base-4 qubit.) 

To derive the ground state consider first the plaquette operator in the $\sigma^z$-basis; clearly a ground state must have an even number of each spin-up and spin-down on every plaquette to be a $+1$ eigenstate of each plaquette operator. Let's consider the state with all spin-up $|\uparrow\uparrow\uparrow\cdots\rangle$; acting with a vertex operator on this flips all the spins around the vertex (see Fig.\ \ref{fig:model}b) giving a state still in 
ground state sector of the plaquette operators as an even number of spins are flipped on the plaquettes surrounding the vertex. (As is also clear from the fact that all the stabilizer operators commute.) The $+1$ eigenstate of that particular vertex operator is thus the symmetric superposition of the two states. A convenient way to express the operation of one or several adjacent vertex operators is in turns of loop traversing the flipped spins. Such loops (fig.\ \ref{fig:model}b-c) generated from products of vertex operators will always be topologically trivial loops on the surface of the torus since they are just constructed by merging the local loop corresponding to a single vertex operator. Successively acting with vertex operators on the states generated from the original  $|\uparrow\uparrow\uparrow\cdots\rangle$ we realize that the ground state is simply the symmetric superposition of all states that are generated from this by acting with (trivial) loops $|\text{GS}_0\rangle=\sum_{i\in \text{all trivial loops}}\text{loop}_i|\uparrow\uparrow\uparrow\cdots\rangle$.

To generate the other ground states we consider the operators $\bar{X}_1$ and $\bar{X}_2$ (Fig.\ \ref{fig:model}d) which are products of $\sigma^x$ corresponding to the two non-trivial loops that wind the torus. (Deformations of these loops just correspond to multiplication by trivial loops and is thus inconsequential.) Correspondingly there are non-trivial loops of $\sigma^z$ operators $\bar{Z}_1$ and $\bar{Z}_2$. The four ground states are thus the topologically distinct states 
$\{|\text{GS}_0\rangle,\bar{X}_1|\text{GS}_0\rangle,\bar{X}_2|\text{GS}_0\rangle,\bar{X}_2\bar{X}_1|\text{GS}_0\rangle\}$ distinguished by their eigenvalues of $\bar{Z}_1$ and $\bar{Z}_2$ being $\pm 1$. For a torus with $d\times d$ plaquettes there are $2d^2$ physical qubits and the code distance, i.e. minimum length of any logical operator ($\bar{X}_i$ or $\bar{Z}_i$), is $d$.  

\subsubsection{Error correction}

Errors in the physical qubits will take the state out of the ground state sector and thereby mask the encoded state. The task of the error correction procedure is to move the system back to the ground state sector without inadvertently performing a logical operation to change the logical qubit state. A $\sigma^x$ error on a physical qubit corresponds to a bit-flip error. On the toric code this gives rise to a pair of defects (a.k.a. quasiparticles or anyons) in the form of neighboring plaquettes with $-1$ eigenvalues of the plaquette stabilizers. Similarly a $\sigma^z$ error corresponds to a phase-flip error which gives rise to a pair of neighboring $-1$ defects on two vertices. A $\sigma^y=i\sigma^x\sigma^z$ simultaneously creates both types of defects.   
A natural error process is to assume that $X,Y,Z$ errors occur with equal probability, so called depolarizing noise. This however requires to treat correlations between $X$ and $Z$ errors and the simpler uncorrelated noise model is often used, which is what we will consider in this work, focusing on bit-flip errors and corresponding plaquette defects. Here $X$ and $Z$ errors occur independently with probability $p$ whereas $Y=XZ$ errors occur with probability $p^2$. Correcting independent $X$ and $Z$ errors is completely equivalent (with defects either on plaquettes or on vertices) and it is therefore sufficient to formulate an error correcting algorithm for one type of error. (For actual realizations of the physical qubits the error process may in fact be intermediate between these two cases\cite{PhysRevLett.120.050505}.) Regardless of noise model and type of error an important aspect of the error correction of a stabilizer formalism is that the entanglement of the logical qubit states or its excitations does not have to be considered explicitly as errors act equivalently on all states that belong to the same stabilizer sector.  

\begin{figure}
    \includegraphics[scale=1.2]{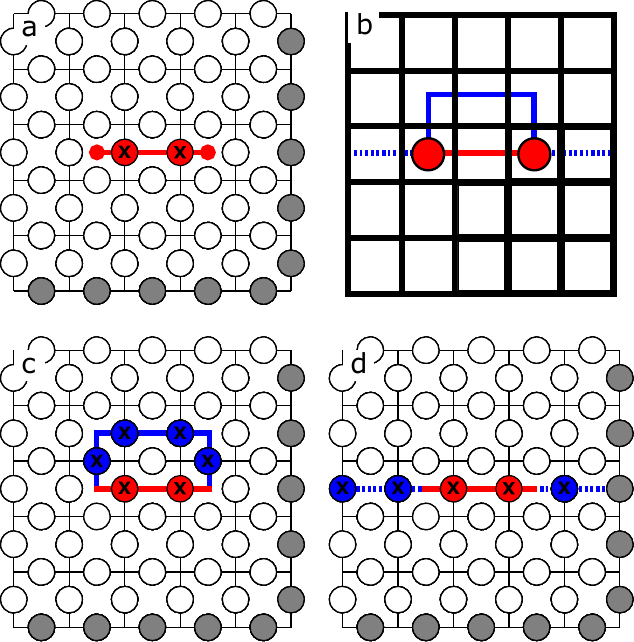} 
    \caption{\label{fig:errorchain} Bit-flip errors (red 'X') and possible error correction bit-flips (blue 'X'). (a) Two neighboring errors and the corresponding error chain (red line) and syndrome (red dots). (b) Visualized in terms of the syndrome  with error chain and two possible correction chains (blue) as expressed explicitly in (c) and (d). The error chain plus the correction chain in (d) constitutes a non-trivial loop and a logical bit-flip operation (as in Figure \ref{fig:model}d), thus a failed error correction, in contrast to the trivial loop in (c).}
\end{figure}

A crucial aspect of quantum error correction is that the actual bit-flip errors cannot be measured without collapsing the state into a partial basis and destroying the qubit. What can be measured without destroying the logical qubit are the stabilizers, i.e. for bit-flip error the parity of the plaquette operators.  The complete set of incorrect ($-1$) plaquettes makes up the syndrome of the state. The complete set of bit-flip errors will produce a unique syndrome as the end-points of strings of bit-flip errors. The converse however is not true, which is what makes the task challenging. In order to do the error correction we need to suggest a number of physical bits that should be flipped in order to achieve the pair-wise annihilation of the defects of the syndrome. Consider a single pair of defects which have been created by a particular chain of errors. (See Figure \ref{fig:errorchain}.) The error correction needs to suggest a correction string connecting the two defects. If this is done properly the correction string and the error string form a trivial loop, thus returning the qubit to the original state. If instead the correction string and the error string together make up a non-trivial loop that winds the torus we have eliminated the error syndrome but changed the state of qubit (corresponding to a logical bit-flip), thus failed the task of correcting the error.   

For the uncorrelated noise model it can be shown, by mapping to the random bond Ising model, that for $d\rightarrow\infty$ there is a critical threshold $p_c\approx 0.11$ below which the most probable correction chains to complement the error chain will with certainty form trivial loops, while for $p>p_c$ non-trivial loops occur with finite probability.\cite{dennis2002topological}  For a finite system, the sharp transition is replaced by a cross-over, as seen in Figure \ref{fig:results}, where for increasing $d$ the fraction of successful error correction evolves progressively towards 1 for $p<p_c$, and to $1/4$ (thus completely unpredictable) for $p>p_c$.   

For the uncorrelated noise model on the torus the most likely set of error chains between pairs of defects which is consistent with a given syndrome would be one that corresponds to the smallest number of total bit flips, i.e. the shortest total error chain length. Thus, a close to optimal algorithm for error correction for this system is the Minimum Weight Perfect Matching (MWPM) algorithm\cite{edmonds1965paths}. (This algorithm is also near optimal for the problem with syndrome errors as long as it is still uncorrelated noise\cite{dennis2002topological,fowler2015minimum}.) The MWPM algorithm for the perfect syndrome corresponds to reducing a fully connected graph, with an even number of nodes and with edges specified by the inter-node distances, to the set of pairs of nodes that minimize the total edge length. This algorithm can be implemented efficiently\cite{kolmogorov2009blossom} and we will use this as the benchmark of our RL results. In fact, as we will see, the RL algorithm that we formulate amounts to solving the MWPM problem. In this sense the work presented in this  paper is to show the viability of the RL approach to this problem with the aim for future generalizations to other problems where MWPM is sub-optimal, such as for depolarizing noise or more general error models. 

\subsection{Q-learning}

Reinforcement learning is a method to solve the problem of finding an optimal policy of an agent acting in a system where the actions of the agent causes transitions between states of the system.\cite{sutton2018reinforcement} The policy $\pi(s,a)$ of an agent describes (probabilistically perhaps) the action $a$ to be taken by the agent when the system is in state $s$. In our case the state will correspond to a syndrome, and an action to moving a defect one step. The optimal policy is the one that gives the agent maximal return (cumulative discounted reward) over the course of its interaction with the system. Reward $r_{t+1}$ is given when the system transitions from state $s_t\rightarrow s_{t+1}$ such that the return starting at time $t$  is given by $R_t=r_{t+1}+\gamma r_{t+2}+\gamma^2 r_{t+3}+\cdots$. Here $\gamma\leq 1$ is the discounting factor that quantifies how we want to value immediate versus subsequent reward. As will be discussed in more detail, in the work presented in this paper a constant reward $r=-1$ will be given for each step taken, so that in practice the optimal policy will be the one that minimizes the number of actions, irrespectively of the value of $\gamma$. (Although in practice, even here the value of $\gamma$ can be important for the convergence of the training.)

One way to represent the cumulative reward depending on a set of actions and corresponding transitions is by means of an action-value function, or Q-function. This function $Q(s,a)$ quantifies the expected return when in state $s$ taking the action $a$, and subsequently following some policy $\pi$. In one-step Q-learning we quantify $Q$ according to $Q(s,a)=r+\gamma\max_{a'}Q(s',a')$, with $s\xrightarrow{a}s'$, which corresponds to following the optimal policy according to our current estimate of $Q$.
In order to learn the value of the Q-function for all states and actions we should explore the full state-action space, with the policy given by taken action $a$ according to $\max_a Q(s,a)$ eventually guaranteed to converge to the optimal policy. 
 However, an unbiased exploration gets prohibitively expensive and it is therefore in general efficient to follow an $\epsilon$-greedy policy which with probability $(1-\epsilon)$ takes the optimal action based on our current estimate of $Q(s,a)$ but with probability $\epsilon$ takes a random action. From what we have learned by this action we would update our estimate for $Q$ according to 
 \begin{equation}
 \label{Q_update}
     Q(s,a)\leftarrow Q(s,a)+\alpha[(r+\gamma\max_{a'}Q(s',a'))-Q(s,a)]\,,
 \end{equation} where $\alpha<1$ is a learning rate. This procedure is then a trade-off between using our current knowledge of the $Q$ function as a guide for the best move to avoid spending extensive time on expensive moves but also exploring to avoid missing out on rewarding parts of the state-action space. 

\subsubsection{Deep Q-learning}

\begin{figure}
       
    \includegraphics[scale=0.45]{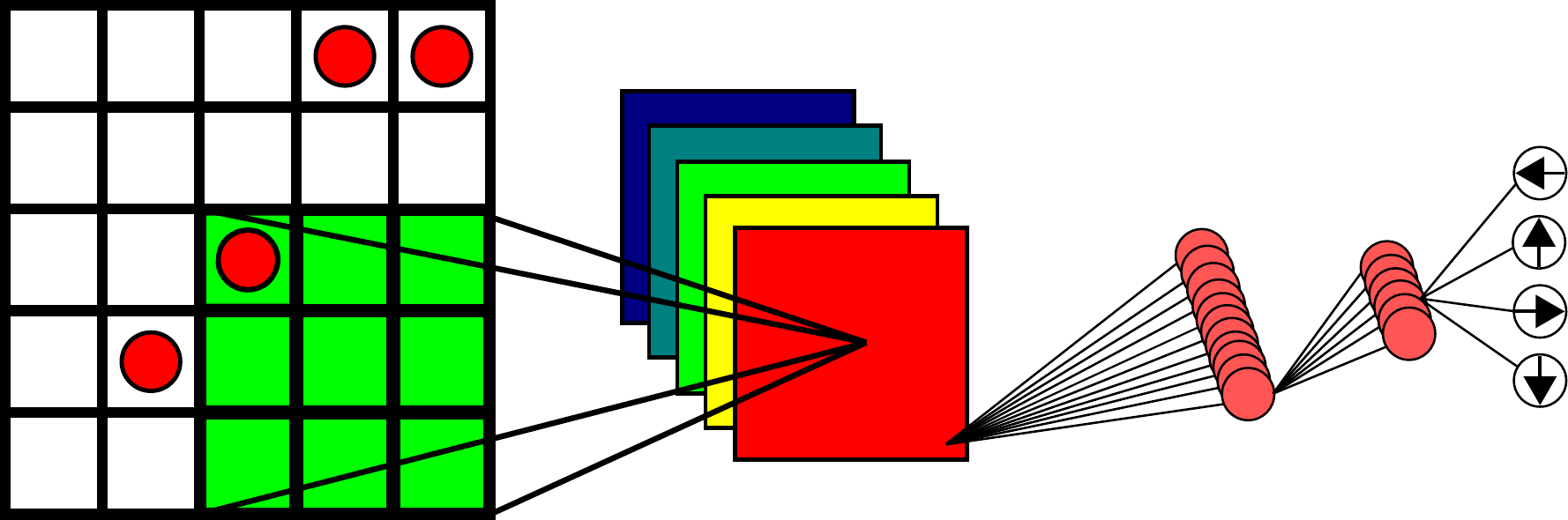} 
    \caption{\label{fig:network}Structure of the deep Q-network. The input layer is a $d\times d$ matrix corresponding to the "perspective" $P$, of one defect of the syndrome. (Using translational symmetry on the torus, any defect can be placed at the center.) The output layer gives the action value $Q(P,a,\theta)$ of moving the central defect to any of the four neighboring plaquettes $a=U,D,R,L$, given the current training state of network parameters $\theta$. The hidden layers consist of a convolutional layer (of which a $3\times 3$ filter is indicated on the input layer) and several fully connected layers. (For details, see Appendix.) Successively scanning all defects using the same network gives the full action value function of the syndrome. }
    
\end{figure}

For a large state-action space it is not possible to store the complete action-value function. (Disregarding symmetries, for a $d\times d$ system with $N_S$ defects, the state space has size $d^2\choose N_S$, $\sim 10^{13}$ for $p\approx 10$\% and $d=7$.) In deep Q-learning\cite{mnih2015human}, the action-value function is instead represented by a deep neural network with the input layer corresponding to some representation of a state and the output layer corresponding to the value of the possible actions. The idea is that similarities in the value of different regions of the state-action space may be stored in an efficient way by the deep network.  
Parametrizing the $Q$-function by means of neural network we write $Q(s,a,\theta)$, where $\theta$ represents the complete set of weights and biases of the network. (We use a convolutional network with $\sim 10^6$ parameters for the $d=7$ problem.) As outlined in more detail in the following sections the latter can be trained using supervised learning based on a scheme similar to one step Q-learning.

\section{RL Algorithm \label{RLsection}}

The decoder presented in this paper is a neural network-based agent optimized using reinforcement learning to observe toric code syndromes and suggesting recovery chains for them step by step. The agent makes use of a deep convolutional neural network, or Q-network, (see Fig. \ref{fig:network}) to approximate Q values of actions given a syndrome.

In a decoding session, a syndrome $S$ corresponding to the coordinates of $N_S$ defects $e_i$ ($i=1,...,N_S$) is fed to the algorithm as input. The syndrome is the state of the system as visible to the agent. The syndrome at any time step is that generated by accumulated actions of the agent on the syndrome given by the initial random distribution of bit-flips. There is also a hidden state corresponding to the joint set of initial and agent flipped qubits. After the complete episode resulting in a terminal state with an empty syndrome, an odd number of non-trivial loops (in either $X_1$ or $X_2$) indicates a failed error correction. In the algorithm used in this work however, the success/fail information does not play any explicit role in the training, except as external verification of the performance of the agent. Instead reward $r=-1$ is given at every step until the terminal state regardless of whether the error correcting string(s) gave rise to an unwanted logical operation. Taking the fewest number of steps to clear the syndrome is thus the explicit target of the agent, corresponding to actuating the MWPM algorithm.  (An alternative formulation with different dependence on $\gamma$ would be to reward $+1$ at the terminal step.) 

It would seem very natural to base the RL reward scheme on the success/failure information from the hidden state. However, we found it difficult to converge to a good agent based on this, for the following reason: given a particular starting syndrome, consistent with a distribution of different error strings, most of these  are properly corrected by the MWPM algorithm whereas a minority are not. As the syndrome is all that the agent sees, it has no chance to learn to distinguish between these two classes, thus trying to use it for training will only obscure the signal. Nevertheless, for future more advanced tasks, such as dealing with noise biased towards bit or phase flips or with spatial variations it will probably be necessary to explore the use of the fail/success information for the reward scheme.


\subsection{State-space formulation}

\begin{figure}
    \includegraphics[scale=.53]{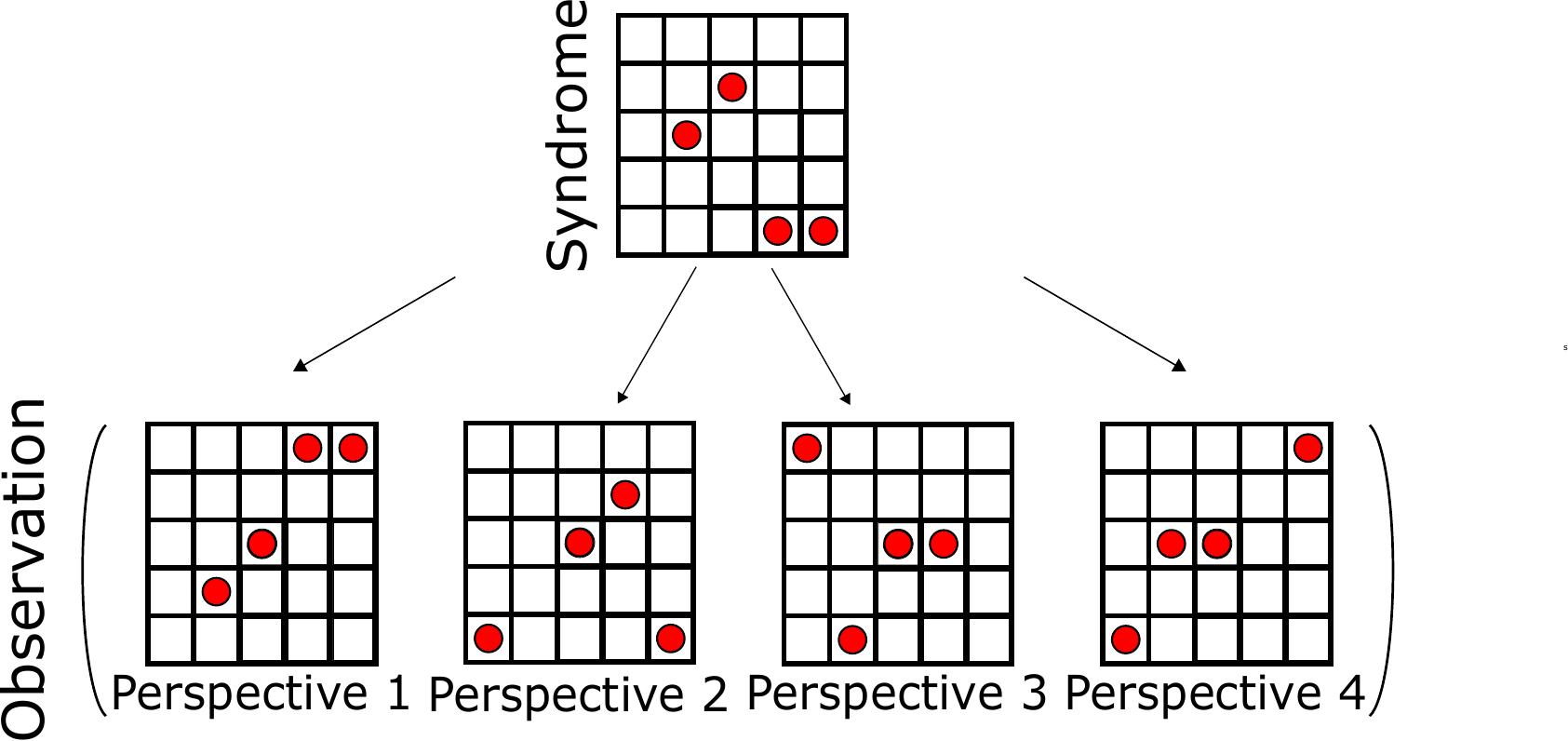} 
    \caption{\label{fig:perspective}State formulation. The toric code syndrome, defines an "observation" that contains the centralized "perspectives" for each defect.}
\end{figure}

Due to the periodic boundary conditions of the code, the syndrome can be represented with an arbitrary plaquette as its center. Centering a defect $e_i$, we define the {\em perspective}, $P_i$, of that defect, consisting of the relative positions of all other defects in the syndrome. The set of all perspectives given a syndrome we define as an {\em observation}, $O$, as exemplified in Figure \ref{fig:perspective}. (The syndrome, observation and perspective all contain equivalent information but represented differently.) 

The agent will be given the option of moving any defect one plaquette in any direction (left, right, up, or down), corresponding to performing a bit flip on one of the physical qubits enclosing the plaquette containing the defect. Clearly the total number of available actions varies with the number of defects, which is inconvenient if we want to represent the Q-function in terms of a neural network. In order for the Q network to have a constant-sized output regardless of how many defects are present in the system, each perspective in the observation is instead sent individually to the Q network. Thus, 
$Q(P,a,\theta)$ represents the value of moving the central defect $a=L,R,U,D$, given the positions of all other defects  specified by the perspective $P$, for network parameters $\theta$. The network with input and output is represented graphically in Figure \ref{fig:network}. The full Q-function corresponding to a syndrome is given by $\{Q(P,a,\theta)\}_{P\in O}$. When the Q value of each action for each defect has been obtained, the choice of action and defect is determined by a greedy policy. The new syndrome is sent to the algorithm and the procedure is repeated until no defects remain.

\subsection{Training the neural network\label{training}}

\begin{figure}
       
    \includegraphics[scale=0.5]{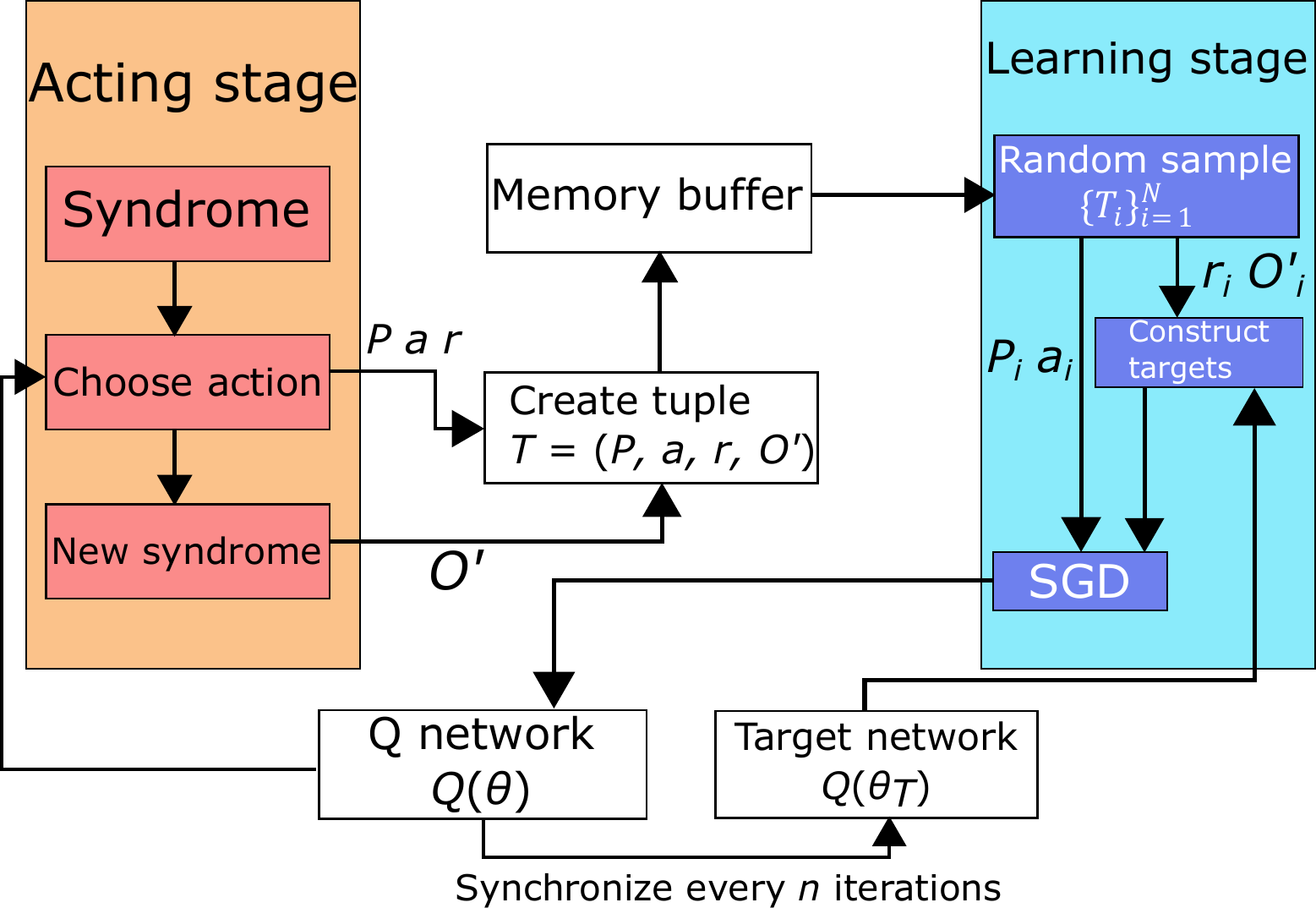} 
    \caption{\label{fig:flowchart2} Flowchart of the training procedure. A learning sequence consists of an acting stage followed by a learning stage. $(P,a,r)$ is (perspective,action,reward) corresponding to what defect is acted on with what action and giving what reward. $O'$ is the observation (see Fig. \ref{fig:perspective}) corresponding to the resulting syndrome. A pseudocode representation is given in Algorithm \ref{alg:our}. (SGD is Stochastic Gradient Descent.)}
    
\end{figure}

Training of the decoder agent was done using the Deep Q Network (DQN) algorithm \cite{mnih2015human}. This algorithm utilizes the technique of experience replay in which the experience acquired by the agent is stored as transition tuples in a memory buffer. When updating the Q network (given by parameters $\theta$), a mini-batch of random samples is drawn from this memory buffer. By taking random samples of experience, the temporal correlation of the data is minimized, resulting in a more stable training procedure of the neural network. To further increase the stability of the training, the DQN algorithm makes use of a target Q network (with parameters $\theta_T$) to compute update targets. The target Q network is periodically synchronized with the updated Q network.

A training sequence begins with an acting stage, where a syndrome is sent to the agent, which uses the Q network $\theta$ to suggest a defect perspective, $P$, and an action, $a$. An $\epsilon$-greedy policy is used by the agent, meaning that it will suggest the action with the highest Q-value with probability $(1-\epsilon)$. Otherwise a random action is suggested. The action is performed on the defect, $e$, corresponding to $P$, resulting in a reward, $r$, and a new observation, $O'$, derived from the resulting syndrome. The whole transition is stored as a tuple, $T = (P, a, r, O')$, in a memory buffer.  After this, the training sequence enters the learning stage using (mini-batch) stochastic gradient descent. First, a random sample of transitions, $\{T_i = (P_i, a_i, r_i, O'_i)\}_{i=1}^N$, of a given batch size, $N$, is drawn with replacement from the memory buffer. (Here the discrete $C_4$ rotational symmetry of the problem is enforced by including all four rotated versions of the same tuple.) The training target value for the Q-network is given by
\begin{equation}
\label{training_target}
    y_i=r_i + \gamma \max_{P'\in O'_i;a'}Q(P',a',\theta_T)\,,
\end{equation}
where $\gamma$ is the discount factor and where the more slowly evolving target network parametrized by $\theta_T$ is used to predict future cumulative award. After this, gradient descent is used to minimize the discrepancy between the targets of the sample and the Q network predictions for it, upgrading the network parameters schematically according to
  $-\nabla_\theta\sum_i(y_i-Q(P_i,a_i,\theta))^2\,$.  
A new training sequence is then started, and with some specified rate, the weights of the target network, $\theta_T$, are synchronized with the Q network $\theta$. A pseudocode description of the procedure is presented in algorithm \ref{alg:our} and an illustration of the different components and procedures of the training algorithm and how they relate to each other is found in Figure \ref{fig:flowchart2}. 

\begin{algorithm}[H]
    \caption{Training the reinforcement learning agent decoder }
    \begin{algorithmic}[1]
        \While{syndrome defects remain}
            \State Get observation $O$ from syndrome \Comment{See figure \ref{fig:perspective}}
            \State Calculate $Q(P,a,\theta)$ using $Q$-network for all perspectives $P\in O$.  
            \State Choose which defect $e$ to move with action $a$ using $\epsilon$-greedy policy
            \State $P \gets $ perspective of defect $e$
            \State Perform action $a$ on defect $e$ 
            \State $r \gets $ reward from taking action $a$ on defect $e$
            \State $O' \gets $ observation corresponding to new syndrome
            \State Store transition tuple $T = (P, a, r, O')$ in memory buffer
            \State Draw a random sample of transition tuples
            \For{each transition tuple $T_i$ in sample}
                \State Construct targets $y_i$ using target network $\theta_T$ and reward $r_i$ according to Eqn. \ref{training_target}.
            \EndFor
            \State Update $Q$-network parameters $\theta$
            \State Every $n$ iterations, synchronize the target network with network, setting $\theta_T=\theta$
        \EndWhile
    \end{algorithmic}
    \label{alg:our}
\end{algorithm}

\section{Result}
Data sets with a fixed error rate of 10\%  were generated to train the agent to operate on a code of a specified size. The syndromes in a data set is fed one at a time to the agent, which operates on it until no errors remain. The data sets also contain information about the physical qubit configuration (the hidden state) of the lattice, which (as discussed in section \ref{RLsection}) is used to check the success rate of the decoder. This is compared to the performance of the MWPM decoder on the same syndromes \cite{kolmogorov2009blossom}. The operation of the trained decoder is similar to the cellular automaton decoders\cite{herold2015cellular,kubica2018cellular} in the sense of providing step by step actions based on the current state of the syndrome. This also means that it could be implemented in parallel with the error generation process by continuously adapting to the introduction of new errors.  

\begin{figure}
    \includegraphics[scale=.32]{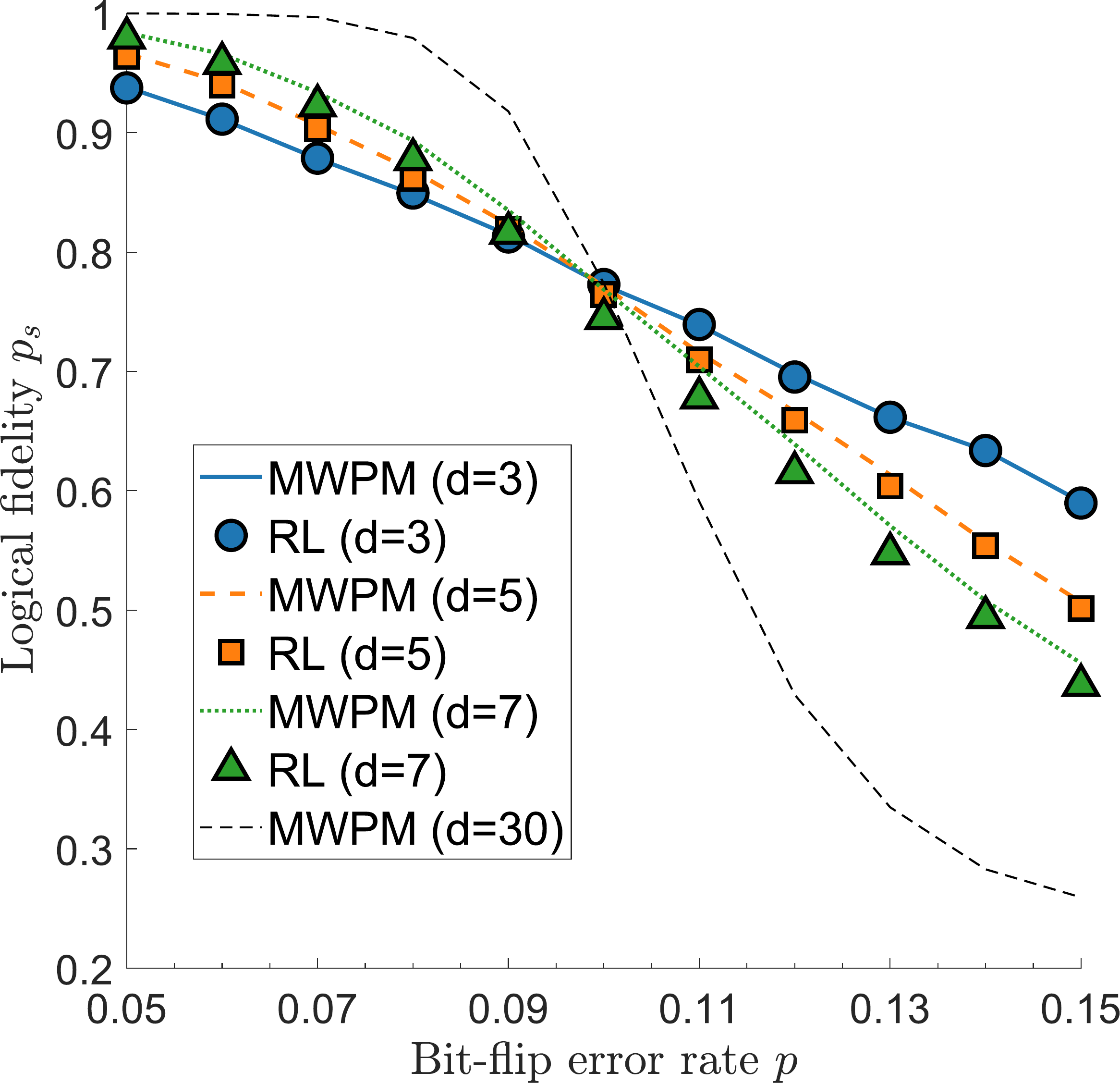}
    \caption{\label{fig:results}Error correction success rate $p_s$ of the converged agents versus bit-flip error rate $p$, for system size $d=3,5,7$, and compared to the corresponding results using MWPM (lines). (The MWPM decoder for $d=30$ is included as a reference for the approach to large $d$.)}
    
\end{figure}

\begin{figure}
    \includegraphics[scale = 0.32]{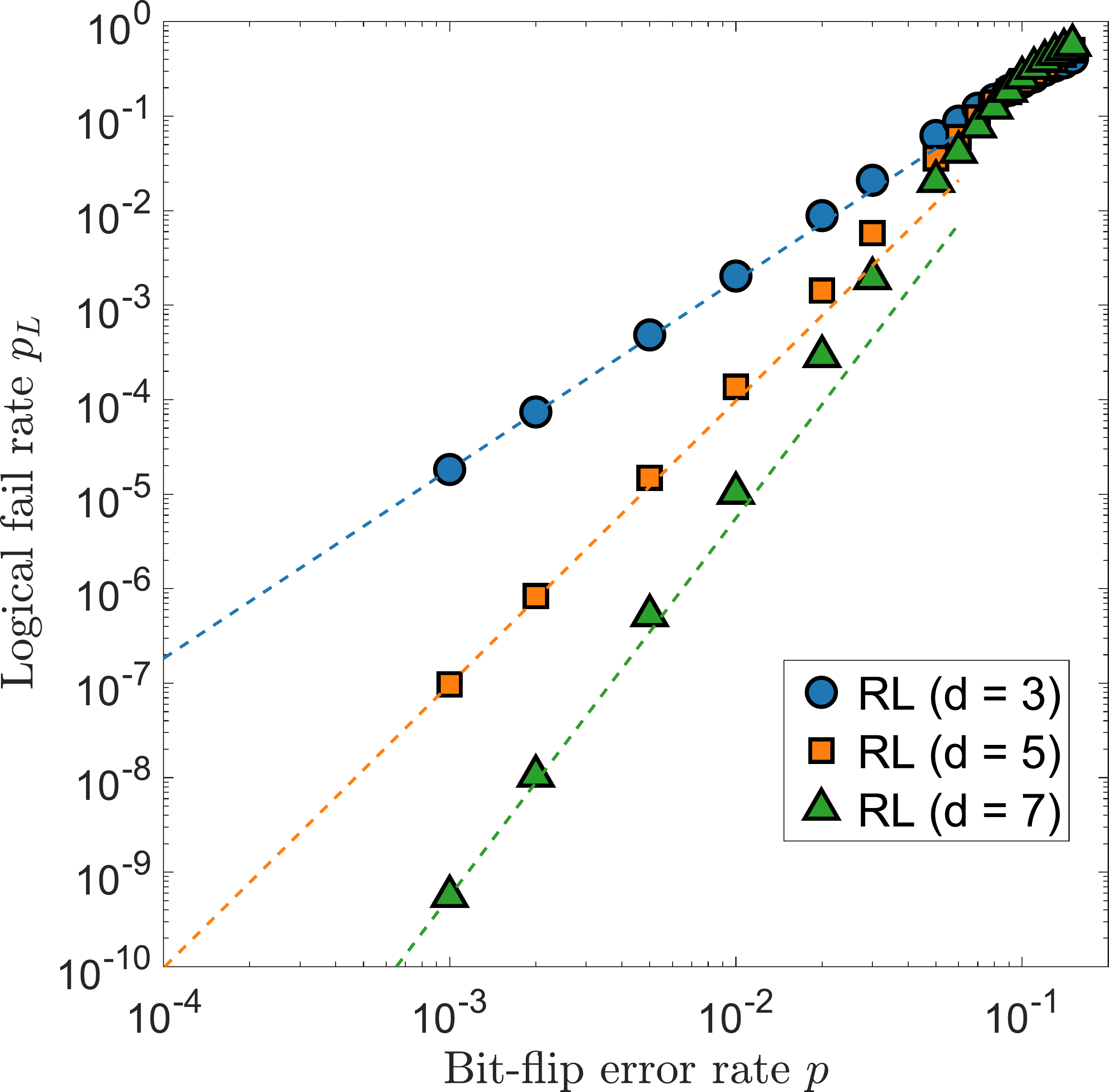}
    \caption{\label{fig:results_smallp}Error correction fail rate $p_L=1-p_s$ shown to converge  to the known asymptotic MWPM behavior (Appendix \ref{app_error}) for small error rates $p\rightarrow 0$. The lines correspond to $p_L\sim p^x$, with $x=\lceil d/2\rceil=2,3,4$ for $d=3,5,7$ fitted to the lowest $p$ data point. }
    
\end{figure}

The proficiency of the well converged agents are shown in figures \ref{fig:results} and \ref{fig:results_smallp} as compared to the MWPM performance.  Given our specified reward scheme, which corresponds to using as few operations as possible, we achieve near optimal results with a performance which is close to that of the MWPM decoder. For small error rates $p_L\rightarrow 0$ it is possible to derive an exact expression for the MWPM fail rate $p_L$ (see Appendix \ref{app_error} and \cite{fowler2012surface,fowler2013optimal}) by explicitly identifying the dominant type of error string. We have checked explicitly that our Q-network agent is equivalent to MWPM for these error strings and thus gives the same asymptotic performance. 

For larger system size $d=9$ we have only been partially successful, with good performance for small error rates, but sub-MWPM performance for larger error rates. Given the exponential growth of the state space this is perhaps not surprising, but by scaling up the hardware and the corresponding size of the manageable Q-network we anticipate that larger code distances would be achievable within the present formalism.

\begin{figure}
    \includegraphics[scale=.9]{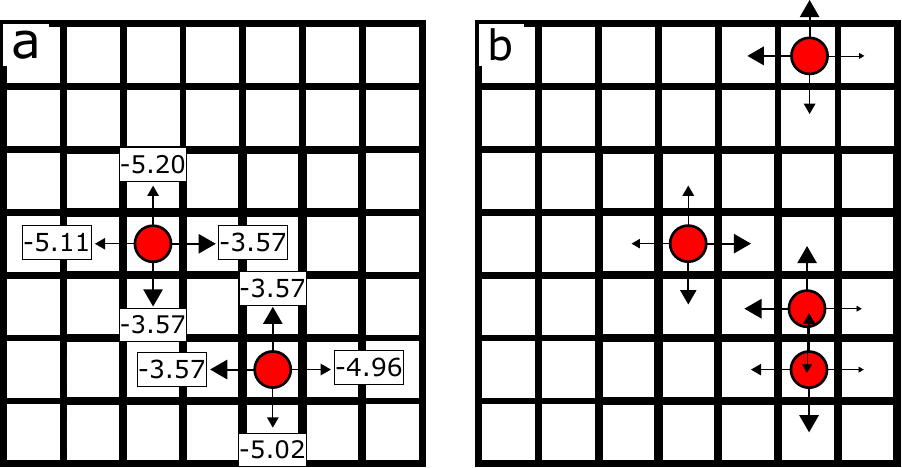}
    \caption{\label{fig:Q_values}Action value function produced by the Q-network for two different syndromes and code distance $d=7$. The magnitude of the arrows indicate the expected return from taking a next step along the arrow and after that following the optimal policy. The optimal policy for the next move corresponds to the action with the biggest arrow(s). In (a) the expected return is written out explicitly, where the best moves are consistent with the constant reward of $-1$ per step and discounting rate $\gamma=0.95$ used in the training.}
\end{figure}

As a demonstration of the operation of the trained agent and the corresponding Q-network we present in Figure \ref{fig:Q_values} the action values $Q(S,a)$ for two different syndromes. (As discussed previously, $Q(S,a)=\{Q(P,a,\theta)\}_{P\in O}$, where $O$ is the observation, or set of perspectives, corresponding to the syndrome $S$.) The size of the arrows are proportional to the discounted return $R$ of moving a defect one initial step in the direction of the arrow and then following the optimal policy. In Fig.\ \ref{fig:Q_values}a, the values are written out explicitly. The best (equivalent) moves have a return $R=-3.57$ which corresponds well to the correct value $R=-1-\gamma-\gamma^2-\gamma^3=-3.62$ for following the optimal policy to annihilate the defects in four steps, with reward $r=-1$ and discount rate $\gamma=.95$. Figure \ref{fig:Q_values}b shows a seemingly challenging syndrome where the fact that the best move does not correspond to annihilating the two neighboring defects is correctly captured by the Q-network. 

One interesting aspect of the close to MWPM performance of the fully trained agent is the ability of the Q-network to suggest good actions independently of how many defects are in the syndrome. A $d=7$ system with $p=10$\% would start out with a syndrome with maybe 20 defects, which is successively pair-wise reduced down to two and finally zero defects, all based on action-values given by the same Q-network ($\theta$). The network is thus surprisingly versatile and capable, given the enormous reduction of the number of adjustable parameters compared to representing and training the full Q-value function as a table.

\section{Conclusions}
In conclusion, we have shown how to implement deep reinforcement learning for quantum error correction on the toric code for moderate size systems using uncorrelated bit-flip (or phase-flip) noise. By training an agent to find the shortest paths for the error correction chains we are able to achieve accuracy close to that using a Minimum Weight Perfect Matching decoder. In order to accomplish this we used the deep Q-network formalism that encodes the action-value function by means of an artificial neural network.\cite{mnih2013playing,mnih2015human} The construction also made good use of the translational invariance on the torus to be able to efficiently reduce the state space representation.

For future work it will be interesting to see how the formalism generalizes to more advanced noise models, imperfect measurements, as well as more general topological codes. Work in progress\cite{Eliasson} indicates that the formalism is in fact readily extended to handle depolarizing noise on the toric code by allowing for the full set of $X$, $Y$, and $Z$ qubit actions. By learning to account for correlations between plaquette and vertex errors super-MWPM performance can be achieved. Also using larger and better adapted convolutional networks allow for somewhat larger system sizes to be addressed. Nevertheless, given the exponential growth of the state-action space it is clear that going much beyond the code distances presented in this paper will require parallelization of the training\cite{horgan2018distributed} as well as massive networks using state of the art hardware, similarly to what is used to achieve super-human performance for board games and computer games.\cite{silver2017mastering,mnih2015human}         

In the longer perspective the main potential of a deep reinforcement learning approach to quantum error correction lies in the fact that is arguably the most promising implementation of AI. Future developments in that area thus opens up also for powerful and flexible machine engineered quantum decoders.

\section*{Acknowledgements}
We thank Niklas Forsstr\"om, Gustav Karlsson, and Elias Hannouch for contributing to the early stages of this work. We also thank Austin Fowler for valuable discussions. Source code can be found at this url: \url{https://github.com/phiandre/ToricCodeRL} 

\bibliographystyle{unsrtnat}
\bibliography{Bib}

\appendix
\section{\label{app_error}Small error rate}
As discussed by Fowler {\em et al.}\cite{fowler2012surface,fowler2013optimal} the likely operating regime of surface code is in the limit of small error rate $p\ll 1$. In addition, in the limit $p\rightarrow 0$ we can derive an exact expression for the rate of logical failure under the assumption of MWPM error correction, thus providing a solid benchmark for our RL algorithm. Such expressions were derived for the surface code in \cite{fowler2013optimal} and here we derive the corresponding expression for bit-flip errors in the toric code. 

Consider first the case of code distance $d$ with $d\in$ odd, which is what we have assumed in the present work. (Using odd $d$ gives an additional simplification of the Q-learning set-up from the fact that any plaquette can be considered the center of the lattice.) As a reminder, the error formulation we use is that every physical qubit has a probability $p$ of bit-flip error, and probability $1-p$ of no error. (In contrast to \cite{fowler2013optimal} we don't consider $\sigma^y$ errors, which would give rise to both bit-flip and phase-flip errors.) For very low $p$, we only need consider states with the minimal number of bit-flip errors that may cause a logical failure. One can readily be convinced (from a few examples) that such states are ones where a number $\lceil d/2\rceil$ (e.g. $\lceil 7/2\rceil=4$) of errors are placed along the path of the shortest possible non-trivial (logical) loops. The latter are $d$ sites long, and on the torus there are $2d$ such loops. For such a state MWPM will always fail, because it will provide a correction string which has $\lfloor d/2\rfloor$ bit-flips rather than the $\lceil d/2\rceil$ flips needed to make a successful error correction. The former correction string, together with the initial error string, will sum to one of the non-trivial (shortest length) loops and give rise to a logical bit-flip. The fail-rate $p_L$, i.e. the fraction of logical fails of all generated syndromes, is thus to lowest order in $p$ and for odd $d$ given by 
\begin{equation}
    p_L=2d\binom{d}{\lceil d/2\rceil}p^{\lceil d/2\rceil}\,.
\end{equation}
Here $2d$ is the number of shortest non-trivial loops, $\binom{d}{\lceil d/2\rceil}$ is the number of ways of placing the errors on such a loop, and $p^{\lceil d/2\rceil}$ is the lowest order term in the probability ($p^{\lceil d/2\rceil}(1-p)^{2d^2-\lceil d/2\rceil}$) of any particular state with $\lceil d/2\rceil$ errors. 

Considering d even (for reference), the corresponding minimal fail scenario has $d/2$ errors on a length $d$ loop. Here the MWPM has a 50\% chance of constructing either a non-trivial or trivial loop, thus giving  the asymptotic fail rate $p_L=d\binom{d}{d/2}p^{d/2}$. 

\section{\label{app}Network architecture and training parameters}

The reinforcement learning agent makes use of a deep convolutional neural network to approximate the Q values for the possible actions of each defect. The network (see Fig. \ref{fig:network}) consists of an input layer which is $d\times d$ matrix corresponding to a perspective (binary input, 0 or 1, with 1 corresponding to a defect), and a convolutional layer followed by several fully-connected layers and an output layer consisting of four neurons, representing each of the four possible actions. All layers have ReLU activation functions except the output layer which has simple linear activation.  
\begin{table}[h]
\caption{\label{tab:d5} Network architecture d=5. FC=Fully connected}
\begin{tabular}{|c|l|l|c|}
\hline
\# & Type & Size & \# parameters\\
\hline
0 & Input  & 5x5  & \\
1 & Conv. &	512 filters; 3x3 size; & \\ 
 & & 2-2 stride & 5 120\\
2 & FC &			256 neurons &  524 544\\
3 & FC &			128 neurons &  32 896\\
4 & FC &			64 neurons &  8 256\\
5 & FC &			32 neurons & 2 080\\
6 & FC (out) &	4 neurons & 132\\
\hline
 & & &  573 028\\
\hline
\end{tabular}
\end{table}
\begin{table}[h]
\caption{\label{tab:d7} Network architecture d=7. }
\begin{tabular}{|c|l|l|c|}
\hline
\# & Type & Size & \# parameters\\
\hline
0 & Input  & 7x7  & \\
1 & Conv. &	512 filters; 3x3 size; & \\ 
 & & 2-2 stride  & 5 120\\
2 & FC &			256 neurons &   1 179 904\\
3 & FC &			128 neurons &  32 896\\
4 & FC &			64 neurons &  8 256\\
5 & FC &			32 neurons & 2 080\\
6 & FC (out) &	4 neurons & 132\\
\hline
 & & &  1 228 388\\
\hline
\end{tabular}
\end{table}

The network architecture is summarized in Table \ref{tab:d5} and \ref{tab:d7}. We also included explicitly a count of the number of parameters (weights and biases) to emphasize the huge reduction compared to tabulating the Q-function. The latter requires of the order $d^2\choose N_S$ entries, for $N_s$ defects, where $N_s$ will also vary as the syndrome is reduced, with initially $N_S\sim 4pd^2$ as each isolated error creates a defect pair and there are $2d^2$ physical qubits.

\begin{figure}
    \includegraphics[scale=.65]{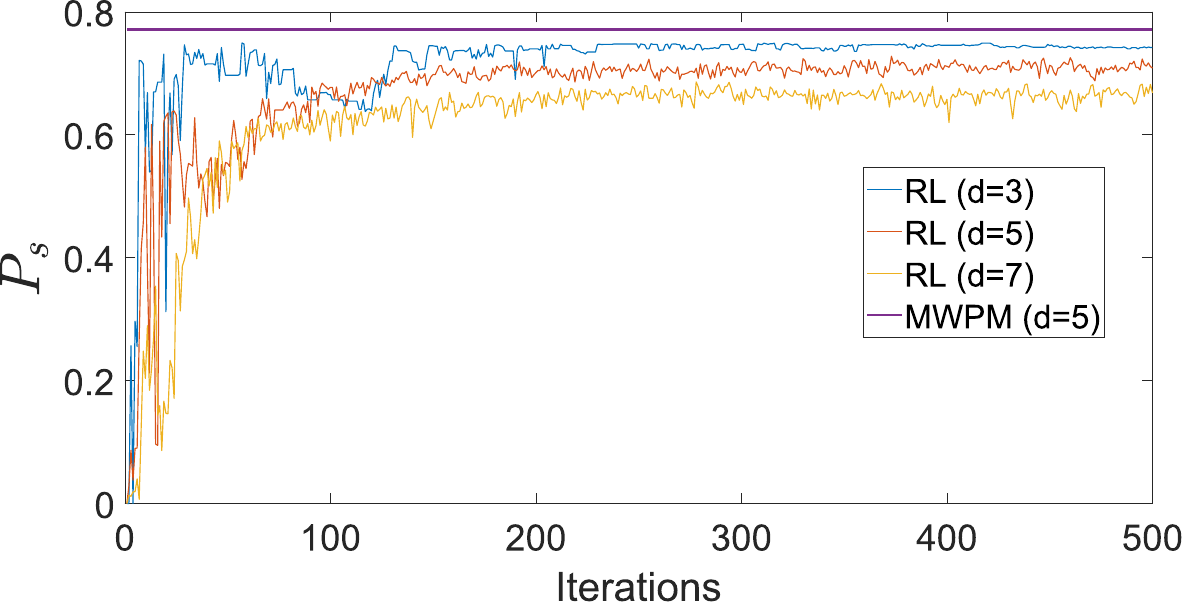}
    \caption{\label{fig:validate}Early training convergence of the Q network agent. Success rate $P_s$ versus number of iterations. One iteration corresponds to annihilating all the defects of a single syndrome. (The very early below $1/4$ success rate is an artifact of using a max count for the number of error correcting steps for the validation.)}
    
\end{figure}

In Figure \ref{fig:validate} we also provide an example of the initial convergence of the algorithm for lattice size $d\times d$, with $d=3,5,7$. Here, each iteration corresponds to solving one syndrome and making the corresponding number of mini-batch training sessions from the experience buffer, as explained in section \ref{training}. A constant set of syndromes is used for the testing so that fluctuations correspond to actual performance variations of the agent.    

In Table \ref{tab:hyper} we list the hyperparameters related to the Q-learning and experience replay set-up, as well as the neural network training algorithm used. The full RL algorithm is coded in Python using Tensorflow and Keras for the Q-network. A single desktop computer was used, with training converging over a matter of hours (for $d=3$) to days (for $d=7$).

\begin{table}[h]
\caption{\label{tab:hyper} Hyperparameters}
\begin{tabular}{|l l|}
\hline
Parameter & Value\\
\hline
discount rate $\gamma$	& 0.95 \\
reward $r$ & -1/step; 0 at finish\\

exploration $\epsilon$ & 0.1\\

max steps  per syndrome & 50\\

mini batch size, $N$ & 32\\

target network update rate       & 100 \\

memory buffer size & 1 000 000\\

optimizer & 'Adam'\\ 

learning rate &	0.001\\
beta$_1$ & 0.9\\
beta$_2$ & 0.999\\
decay & 0.0\\
\hline
\end{tabular}
\end{table}



\end{document}